\documentclass[showpacs,prb,twocolumn]{revtex4}
\usepackage{mathrsfs}
\usepackage{amsfonts}
\usepackage{amsmath}
\usepackage{amssymb}
\usepackage{graphicx}
\usepackage{txfonts}%
\providecommand{\U}[1]{\protect\rule{.1in}{.1in}}

\begin{document}
\title{Quantum many-body theory of qubit decoherence in a finite-size spin bath. II.
Ensemble dynamics}
\author{Wen Yang}
\author{Ren-Bao Liu}
\thanks{rbliu@phy.cuhk.edu.hk}
\affiliation{Department of Physics, The Chinese University of
Hong Kong, Shatin, N. T., Hong Kong, China}


\begin{abstract}
Decoherence of a center spin or qubit in a spin bath is essentially
determined by the many-body bath evolution. The bath dynamics can
start either from a pure state or, more generally, from a
statistical ensemble. In the preceding article [W. Yang and R. B.
Liu, Phys. Rev. B \textbf{78}, 085315 (2008)], we have developed the
cluster-correlation expansion (CCE) theory for the so-called
single-sample bath dynamics initiated from a factorizable pure
state. Here we present the ensemble CCE theory, which is based on
similar ideas of the single-sample CCE: The bath evolution is
factorized into the product of all possible cluster correlations,
each of which accounts for the authentic (non-factorizable)
collective excitation of a group of bath spins, and for the
finite-time evolution in the qubit decoherence problem, convergent
results can be obtained by truncating the ensemble CCE by keeping
cluster correlations up to a certain size. A difference between the
ensemble CCE and single-sample CCE is that the mean-field treatment
in the latter formalism of the diagonal part of the spin-spin
interaction in the bath is not possible in the former case. The
ensemble CCE can be applied to non-factorizable initial states. The
ensemble CCE is checked against the exact solution of an XY spin
bath model. For small spin baths, it is shown that single-sample
dynamics is sensitive to the sampling of the initial state from a
thermal ensemble and hence very different from the ensemble average.
\end{abstract}
\maketitle

\section{Introduction}

The dissipative dynamics of a center spin in a spin bath\cite{Prokofev2000_RPP}
is an old topic in spin resonance spectroscopy.\cite{Kubo1954_JPSJ,Pines1955,Feher1959,Klauder1962}
Recently, this subject is revisited\cite{Fujisawa2002,Elzerman2004_Nature,Kroutvar2004_Nature,Khaetskii2000,
Woods2002,Golovach2004,Semenov2004,Witzel2005,Witzel2006,Witzel2007,Witzel2007_PRB,Witzel2007_PRBCDD,
Yao2006_PRB,Yao2007_RestoreCoherence,Liu2007_NJP,Deng2006,Saikin2007,YangDQD2008,Yang2008CCE}
mostly due to the decoherence issue in quantum information processing.\cite{Loss1998,Imamoglu1999,Awschalom2002_book}
Being a most promising candidate for solid state qubits, electron spins in quantum dots or
impurity centers experience decoherence by coupling to complex solid-state environments.
A series of theoretical\cite{Khaetskii2000,Woods2002,Golovach2004,Semenov2004} and
experimental\cite{Fujisawa2002,Elzerman2004_Nature,Kroutvar2004_Nature} works have
identified that the dominating decoherence mechanism for electron spin qubits at low temperatures
(such as below a few Kelvins) is the entanglement with nuclear spins of the
host lattice.\cite{Witzel2005,Witzel2006,Witzel2007,Witzel2007_PRB,Witzel2007_PRBCDD,
Yao2006_PRB,Yao2007_RestoreCoherence,Liu2007_NJP,Deng2006}

When the qubit flip is suppressed (usually by the large Zeeman energy mismatch
between qubit and bath spins in a moderate magnetic field), the Hamiltonian for
a qubit-bath system has the general form
\begin{equation}
\hat{H}=\left\vert +\right\rangle \hat{H}^{(+)}\left\langle +\right\vert
+\left\vert -\right\rangle \hat{H}^{(-)}\left\langle -\right\vert.
\label{pure_dephasing_H}%
\end{equation}
The bath dynamics is driven by different Hamiltonians $\hat{H}^{(\pm)}$
depending on the qubit states $\left\vert \pm\right\rangle $. For a given
initial bath state $\left\vert \mathcal{J}\right\rangle $ (which could be
a random sampling from a thermal ensemble), the qubit coherence
at time $T$ is characterized by the ``single-sample'' propagator $\langle
\mathcal{J}|e^{i\hat{H}^{(-)}{t}}e^{-i\hat{H}^{(+)}{t}}|\mathcal{J}\rangle$.
For a thermal ensemble of bath states characterized by a density matrix
$\hat{\rho}$, a further ensemble average should be processed and the qubit
coherence is given by the ensemble average
$\operatorname*{Tr}\left[\hat{\rho}e^{i\hat{H}^{(-)}{t}}e^{-i\hat{H}^{(+)}{t}}\right]$.
In general, the key is to evaluate the ensemble-averaged propagator
\begin{equation}
\mathcal{L}=\operatorname*{Tr}\left(  \hat{\rho}e^{i\hat{O}^{(1)}}e^{i\hat
{O}^{(2)}}\cdots\right)  \label{L_ENS}%
\end{equation}
for a general density matrix $\hat{\rho}$ and
arbitrary bath interaction operators $\{\hat{O}^{(j)}\}$.
To address this problem, a
variety of quantum many-body theories have been developed, including the
density matrix cluster expansion,\cite{Witzel2005,Witzel2006,Witzel2007,Witzel2007_PRB,Witzel2007_PRBCDD} the
pair-correlation approximation,\cite{Yao2006_PRB,Yao2007_RestoreCoherence,Liu2007_NJP} and the
linked-cluster expansion.\cite{Saikin2007} The pair-correlation approximation
provides a clear physical picture for the bath dynamics by keeping only
spin-pair correlations. The linked-cluster expansion accurately takes into account
higher-order correlations with a Feynman diagram method, which, however,
becomes dramatically tedious with increasing the order of diagrams.
The density matrix cluster expansion simplifies the evaluation of
higher-order correlations, but it may not converge to the exact results
for relatively small baths.\cite{Yang2008CCE}

Very recently, we have developed a cluster-correlation expansion
(CCE) theory\cite{Yang2008CCE} for the evaluation of the
single-sample
propagator $L_{\mathcal{J}}\equiv\langle\mathcal{J}|e^{i\hat{O}^{(1)}}e^{i\hat{O}^{(2)}%
}\cdots|\mathcal{J}\rangle$, which is a special case of the
ensemble-averaged propagator in Eq.~(\ref{L_ENS}). The CCE method
provides a simple and accurate method to systematically take into
account the high-order correlations. For a temperature $T$ much
higher than the bath interaction strength ($\sim$10$^{-9}$~K for
nuclear spins in GaAs), the initial thermal ensemble can be well
approximated as $\hat{\rho}\approx \exp(-\hat{H}_0/k_BT)$, where
$\hat{H}_0$ is the non-interacting Hamiltonian containing only the
Zeeman energy. Such an initial ensemble is factorizable and a
sampling $\left\vert\mathcal{J}\right\rangle $ from the ensemble can
be taken as a product state $\left\vert \mathcal{J}\right\rangle
=\bigotimes_{n}\left\vert j_{n}\right\rangle$ of all constituent
bath spins, where $|j_{n}\rangle$ denotes the Zeeman energy eigen
state of the $n$th bath spin. For a large spin bath, previous
study\cite{Liu2007_NJP} has shown that the qubit decoherence is
insensitive to the random sampling of the initial bath state from a
thermal ensemble since the statistical fluctuation scales with the
number of bath spins $N$ as $1/\sqrt{N}$. Thus the ensemble dynamics
can be just identified with the single-sample dynamics with a random
choice of the initial state.\cite{Liu2007_NJP} For a relatively
small bath, however, the single-sample dynamics could be sensitive
to the sampling of the initial state and the ensemble average can be
very different from any single sample. More importantly, if the
initial state of the bath is entangled, i.e.,
\begin{equation}
\hat{\rho} \ne \sum_{\alpha}P_{\alpha}\bigotimes_{i} \hat{\rho}^{(\alpha)}_{\{i\}},
\end{equation}
for any choice of probability distribution $\{P_{\alpha}\}$ and
single spin density matrices $\{\hat{\rho}^{(\alpha)}_{\{i\}}\}$,
the single-sample CCE is not applicable.
To extend to general ensemble bath dynamics, one could simply
use the Monte Carlo simulation with a sufficiently large random sampling
of the initial states from the ensemble. The Monte Carlo
simulation is practically cumbersome due to the large
number of initial states required for a faithful reproduction of the
ensemble dynamics, and more importantly, it cannot be applied to non-factorizable
initial states. In this paper we will develop a CCE formalism suitable
for direct evaluation of ensemble-averaged bath evolution.

In Sec. II, we will present the ensemble CCE and compare it with
the single-sample CCE. In Sec. III, we check the ensemble CCE
against the exact solution of a one-dimensional XY mode and
compare the single-sample CCE and ensemble CCE.
Sec. IV gives the conclusions.

\section{Ensemble Cluster-correlation expansion}

\subsection{An example}
\label{example}

Let us consider a bath consisting of $N$ spins and
evaluate the bath evolution
\begin{equation}
\mathcal{L}\equiv\operatorname*{Tr}\left(  \hat{\rho}e^{i\hat{O}}\right),
\label{L_ENS_SINGLE}%
\end{equation}
averaged over a noninteracting (factorizable) ensemble
\begin{equation}
\hat{\rho}=\hat{\rho}_{\{1\}}\otimes\hat{\rho}_{\{2\}}\otimes\cdots\otimes\hat{\rho}_{\{N\}},
\label{factorized}
\end{equation}
where $\hat{\rho}_{\{i\}}=\sum_jp_j|j\rangle\langle j|$ is the
non-interacting density matrix for the $i$th spin $\hat{\mathbf
J}_i$ and
\begin{equation}
\hat{O}\equiv\sum_{n}\alpha_{n}\hat{J}_{n}^{z}+\sum_{m<n}\beta_{m,n}\left(
\hat{J}_{m}^{+}\hat{J}_{n}^{-}+\hat{J}_{m}^{-}\hat{J}_{n}^{+}\right)
\label{OP_A}%
\end{equation}
is the dimensionless bath interaction operator.
Here $\beta_{m,n}$ is the interaction strength between spins $m$ and $n$.
The coefficients $\{\beta_{m,n}\}$ are treated as small quantities.
In the absence of interaction $(\{\beta_{m,n}\}=0)$, the propagator
assumes a factorized form
\[
\left.  \mathcal{L}\right\vert _{\{\beta_{m,n}\}=0}=\mathcal{L}_{\{1\}}%
\mathcal{L}_{\{2\}}\cdots\mathcal{L}_{\{N\}},
\]
where $\mathcal{L}_{\{n\}}\equiv\tilde{\mathcal{L}}_{\{n\}}\equiv
\operatorname*{Tr}(\hat{\rho}_{\{n\}}e^{i\hat{O}_{\{n\}}})$ and
$\hat {O}_{\{n\}}\equiv\alpha_{n}\hat{J}_{n}^{z}$. For
$\{\beta_{m,n}\}\ne 0$, we introduce additional factors (cluster
correlations) to account for the interaction corrections. These
cluster correlations can be introduced successively as follows.
\begin{enumerate}
\item Two-spin correlations $\{\mathcal{\tilde{L}}_{\{i,j\}}\}$.\newline
If the bath consists of only two spins with indices $\{i,j\}$, the propagator would be
\[
\mathcal{L}_{\{i,j\}}\equiv\operatorname*{Tr}\left(  \hat{\rho}_{\{i,j\}}%
e^{i\hat{O}_{\{i,j\}}}\right)
\]
with $\hat{\rho}_{\{i,j\}}\equiv\hat{\rho}_{\{i\}}\otimes\hat{\rho}_{\{j\}}$ and%
\[
\hat{O}_{\{i,j\}}\equiv\alpha_{i}\hat{J}_{i}^{z}+\alpha_{j}\hat{J}_{j}%
^{z}+\beta_{i,j}\left(  \hat{J}_{i}^{+}\hat{J}_{j}^{-}+\hat{J}_{i}^{-}\hat
{J}_{j}^{+}\right)  ,
\]
i.e., $\mathcal{L}_{\{i,j\}}$ is obtained from Eq. (\ref{L_ENS_SINGLE}) by
dropping all spins except $i$ and $j$. Without interaction
($\beta_{i,j}=0$), the propagator is
\[
\left.  \mathcal{L}_{\{i,j\}}\right\vert _{\beta_{i,j}=0}=\mathcal{L}%
_{\{i\}}\mathcal{L}_{\{j\}}=\tilde{\mathcal{L}}_{\{i\}}\tilde{\mathcal{L}%
}_{\{j\}}.
\]
The interaction correction makes the factorization to be
$\mathcal{L}_{\{i,j\}}=\tilde{\mathcal{L}}_{\{i\}}\tilde{\mathcal{L}}%
_{\{j\}}\tilde{\mathcal{L}}_{\{i,j\}}$. Thus the two-spin correlation
is defined as
\begin{equation}
\tilde{\mathcal{L}}_{\{i,j\}}\equiv\frac{\mathcal{L}_{\{i,j\}}}{\tilde
{\mathcal{L}}_{\{i\}}\tilde{\mathcal{L}}_{\{j\}}}. \label{L_IJ_SINGLE}%
\end{equation}
Obviously, the Taylor expansion of the pair correlation with respect
to the interaction strength is
\begin{equation}
\ln\tilde{\mathcal{L}}_{\{i,j\}}=c_{1}\beta_{i,j}+c_{2}\beta_{ij}^{2}%
+\cdots=O(\beta), \label{CE_IJ}%
\end{equation}
where $\beta$ denotes the typical magnitude of the interaction
strength $\{\beta_{m,n}\}$. Thus $\ln\tilde{\mathcal{L}}_{\{i,j\}}$
is at most a first-order small quantity.

\item Three-spin correlations $\{\mathcal{\tilde{L}}_{\{i,j,k\}}\}$.\newline
For a bath of three spins $\{i,j,k\}$, the propagator is
\[
\mathcal{L}_{\{i,j,k\}}\equiv\operatorname*{Tr}\left(  \hat{\rho}%
_{\{i,j,k\}}e^{i\hat{O}_{\{i,j,k\}}}\right)
\]
with $\hat{\rho}_{\{i,j,k\}}\equiv\hat{\rho}_{\{i\}}\otimes\hat{\rho}_{\{j\}}\otimes
\hat{\rho}_{\{k\}}$ and
\[
\hat{O}_{\{i,j,k\}}\equiv\sum_{n=i,j,k}\alpha_{n}\hat{J}_{n}^{z}%
+\sum\limits_{m,n=i,j,k}^{m<n}\beta_{m,n}\left(  \hat{J}_{m}^{+}\hat{J}%
_{n}^{-}+\hat{J}_{m}^{-}\hat{J}_{n}^{+}\right),
\]
i.e., $\mathcal{L}_{\{i,j,k\}}$ is obtained from Eq.~(\ref{L_ENS_SINGLE}) by
dropping all spins except $i$, $j$, and $k$. Similar to the two-spin case,
$\mathcal{L}_{\{i,j,k\}}$ can be factorized as
\[
\mathcal{L}_{\{i,j,k\}}=\tilde{\mathcal{L}}_{\{i\}}\tilde{\mathcal{L}}%
_{\{j\}}\tilde{\mathcal{L}}_{\{k\}}\tilde{\mathcal{L}}_{\{i,j\}}%
\tilde{\mathcal{L}}_{\{j,k\}}\tilde{\mathcal{L}}_{\{i,k\}}\tilde{\mathcal{L}%
}_{\{i,j,k\}},
\]
where
\begin{equation}
\tilde{\mathcal{L}}_{\{i,j,k\}}\equiv\frac{\mathcal{L}_{\{i,j,k\}}}%
{\tilde{\mathcal{L}}_{\{i\}}\tilde{\mathcal{L}}_{\{j\}}\tilde{\mathcal{L}%
}_{\{k\}}\tilde{\mathcal{L}}_{\{i,j\}}\tilde{\mathcal{L}}_{\{j,k\}}%
\tilde{\mathcal{L}}_{\{i,k\}}}, \label{L_IJK_SINGLE}%
\end{equation}
accounts for the non-factorizable correlation among the three spins.
$\ln\tilde{\mathcal{L}}_{\{i,j,k\}}$ would vanish if the interactions
in $\hat{O}_{\{i,j,k\}}$ cannot connect the three spins
$\{i,j,k\}$ into a linked cluster. For example, if $\beta_{i,j}=\beta_{i,k}=0$
and $\beta_{j,k}\ne 0$, the three spin propagator
$\mathcal{L}_{\{i,j,k\}}$ would be factorized as
\[
\mathcal{L}_{\{i,j,k\}}=\mathcal{L}_{\{i\}}\mathcal{L}_{\{j,k\}},
\]
which, together with $\ln\tilde{\mathcal{L}}_{\{i,j\}}=\ln\tilde{\mathcal{L}%
}_{\{i,k\}}=0$ [according to Eq.~(\ref{CE_IJ})], leads to $\ln\tilde
{\mathcal{L}}_{\{i,j,k\}}=0$ according to Eq.~(\ref{L_IJK_SINGLE}). This
connectivity property of $\ln\tilde{\mathcal{L}}_{\{i,j,k\}}$ leads to the
Taylor expansion%
\begin{equation}
\ln\tilde{\mathcal{L}}_{\{i,j,k\}}=c_{1}\beta_{i,j}\beta_{i,k}+c_{2}%
\beta_{j,i}\beta_{j,k}+c_{3}\beta_{k,i}\beta_{k,j}+O(\beta^{3}).
\label{CE_IJK}
\end{equation}
Thus $\ln\tilde{\mathcal{L}}_{\{i,j,k\}}$ is at most a second-order
small quantity.

\item Cluster correlation $\{\mathcal{\tilde{L}}_{\mathcal{C}}\}$.\newline
The above factorization procedure can be carried out for baths consisting of
more and more spins. For a bath of an arbitrary group of $n$ spins (denoted as $\mathcal{C}$),
the propagator becomes
\[
\mathcal{L}_{\mathcal{C}}\equiv\operatorname*{Tr}\left(  \hat{\rho
}_{\mathcal{C}}e^{i\hat{O}_{\mathcal{C}}}\right)  ,
\]
which is obtained from Eq.~(\ref{L_ENS_SINGLE}) by dropping all spins except
those belonging to the group $\mathcal{C}$. By introducing the cluster correlation
\[
\tilde{\mathcal{L}}_{\mathcal{C}}\equiv\frac{\mathcal{L}_{\mathcal{C}}}%
{\prod\limits_{\mathcal{C}^{\prime}\subset\mathcal{C}}\tilde{\mathcal{L}%
}_{\mathcal{C}^{\prime}}},
\]
the propagator is factorized as
\[
\mathcal{L}_{\mathcal{C}}=\prod_{\mathcal{C}^{\prime}\subseteq\mathcal{C}%
}\tilde{\mathcal{L}}_{\mathcal{C}^{\prime}}.
\]
By mathematical induction, it can be readily proved that $\ln\tilde
{\mathcal{L}}_{\mathcal{C}}$ vanishes if the interactions contained
in $\hat{O}_{\mathcal{C}}$ cannot connect all the spins in group
$\mathcal{C}$ into a linked cluster. Such connectivity property
ensures that in each term of the Taylor expansion of
$\ln\tilde{\mathcal{L}}_{\mathcal{C}}$ about the interaction
strength, the coefficients $\{\beta_{i,j}\}$'s must appear at least
$(\left\vert \mathcal{C}\right\vert -1)$ times ($\left\vert
\mathcal{C}\right\vert $ being the number of spins in the group).
Thus $\ln\tilde{\mathcal{L}}_{\mathcal{C}}=O\left(\beta^{\left\vert
\mathcal{C}\right\vert -1}\right)$.
\end{enumerate}
In particular, the full propagator $\mathcal{L}$ of the whole bath is factorized into the
product of all possible cluster correlations as
\begin{equation}
\mathcal{L}=\left(  \prod_{i}\tilde{\mathcal{L}}_{\{i\}}\right)  \left(
\prod_{\{i,j\}}\tilde{\mathcal{L}}_{\{i,j\}}\right)  \cdots\tilde{\mathcal{L}%
}_{\{1,2,\cdots,N\}}=\prod_{\mathcal{C}\subseteq\{1,2,\cdots,N\}}%
\tilde{\mathcal{L}}_{\mathcal{C}}. \label{CCE_ENS_SINGLE}%
\end{equation}

An exact evaluation of the ensemble CCE in Eq.
(\ref{CCE_ENS_SINGLE}) is not possible in general, which amounts to
exactly solving the many-spin dynamics. In the qubit decoherence
problem, it often suffices to truncate the CCE to an appropriate
order $M$ (denoted as $M$-CCE for short) by dropping all cluster
correlations with sizes larger than $M$,
\begin{equation}
{\mathcal{L}}^{\left(  M\right)  }=\prod_{\left\vert \mathcal{C}\right\vert
\leq M}\tilde{\mathcal{L}}_{\mathcal{C}}. \label{CCEM_ENS_SINGLE}%
\end{equation}
For example, the first-order truncation of the ensemble CCE (the 1-CCE) is
\begin{equation}
\mathcal{L}^{(1)}=\tilde{\mathcal{L}}_{\{1\}}\tilde{\mathcal{L}}_{\{2\}}%
\cdots\tilde{\mathcal{L}}_{\{N\}}=\prod_{i}\tilde{\mathcal{L}}_{\{i\}},
\label{CCE1_ENS_SINGLE}%
\end{equation}
which is equivalent to Eq. (\ref{L_ENS_SINGLE}) with all interaction terms in
$\hat{O}$ dropped. In order to incorporate the interaction effects, the lowest
nontrivial order of truncation is the second order (2-CCE),
\begin{equation}
{\mathcal{L}}^{\left(  2\right)  }=\left(  \prod_{i}\tilde{\mathcal{L}%
}_{\{i\}}\right)  \left(  \prod_{\{i,j\}}\tilde{\mathcal{L}}_{\{i,j\}}\right)
, \label{CCE2_ENS_SINGLE}%
\end{equation}
which coincides with the pair-correlation approximation.\cite{Yao2006_PRB,Liu2007_NJP}

Since only connected clusters for which $\ln\tilde{\mathcal{L}}_{\mathcal{C}
}\neq0$ contribute to the propagator $\mathcal{L}$, the convergence (and
hence the justification for the truncation) of the ensemble CCE can be
estimated as follows. First, if each spin interacts, on average, with $q$
spins, then the number of connected size-$M$ clusters is $\sim Nq^{M-1}$, with
$N$ the total number of bath spins. Second, for a size-$M$ cluster,
$\ln\tilde{\mathcal{L}}_{\mathcal{C}}=O\left(\beta^{M-1}\right)$.
The contribution to $\ln{\mathcal{L}}$ from all the size-$M$ clusters is $\sum_{\left\vert \mathcal{C}
\right\vert =M}\ln\tilde{\mathcal{L}}_{\mathcal{C}}\sim N(q\beta)^{M-1}$.
For $q\beta<1,$ the ensemble CCE converges.

\subsection{General theory}

The above example can be readily generalized to
\begin{equation}
\mathcal{L}=\operatorname*{Tr}\left(\hat{\rho}e^{i\hat{O}^{(1)}}e^{i\hat
{O}^{(2)}}\cdots\right),  \label{L_ENS_MULTI}%
\end{equation}
with a general ensemble $\hat{\rho}$ and an arbitrary
series of time-ordered bath operators $\hat{O}^{(j)}$ ($j=1,2,\cdots$).
The bath interactions need not be purely off-diagonal or contain only pairwise interactions,
and the initial density matrix need not be factorizable.
For a thermal ensemble as
\begin{equation}
\hat{\rho}=\exp\left(-\beta_T \hat{H}\right),
\end{equation}
the density matrix itself can be viewed as a propagator with imaginary time $\tau=-i\beta_T$
and the whole propagator can be written as
\begin{equation}
\mathcal{L}=\operatorname*{Tr}\left(\hat{\rho}_{0}e^{i\left(i\beta_T\hat{H}\right)}e^{i\hat{O}^{(1)}}e^{i\hat
{O}^{(2)}}\cdots\right),
\end{equation}
with $\hat{\rho}_{0}$ denoting the trivial thermal state at infinite temperature ($\beta_T=0$).

In essentially the same way as illustrated in the example above, a hierarchy of
cluster correlations $\{\tilde{\mathcal{L}}_{\mathcal{C}}\}$
can be introduced.
First, the single-spin correlation is defined as
\[
\tilde{\mathcal{L}}_{\{i\}}\equiv\mathcal{L}_{\{i\}},
\]
where
\[
\mathcal{L}_{\{i\}}\equiv\operatorname*{Tr}\left(  \hat{\rho}_{\{i\}}%
e^{i\hat{O}_{\{i\}}^{(1)}}e^{i\hat{O}_{\{i\}}^{(2)}}\cdots\right)
\]
is obtained from Eq.~(\ref{L_ENS_MULTI}) by dropping all spins except spin
$i$ and
\begin{equation}
\hat{\rho}_{\{i\}}\equiv \text{Tr}_{k\ne i}\left[\hat{\rho}\right],
\end{equation}
is the reduced density matrix of the $i$th spin.
Then the cluster correlation for an arbitrary group $\mathcal{C}$ of
bath spins is defined as
\[
\tilde{\mathcal{L}}_{\mathcal{C}}\equiv\frac{\mathcal{L}_{\mathcal{C}}}%
{\prod\limits_{\mathcal{C}^{\prime}\subset\mathcal{C}}\tilde{\mathcal{L}%
}_{\mathcal{C}^{\prime}}},
\]
where
\[
\mathcal{L}_{\mathcal{C}}\equiv\operatorname*{Tr}\left(  \hat{\rho
}_{\mathcal{C}}e^{i\hat{O}_{\mathcal{C}}^{(1)}}e^{i\hat{O}_{\mathcal{C}}^{(2)}%
}\cdots\right),
\]
is obtained from Eq.~(\ref{L_ENS_MULTI}) by dropping all bath spins outside
group $\mathcal{C}$ and
\begin{equation}
\hat{\rho}_{\mathcal C}\equiv \text{Tr}_{k\notin{\mathcal C}}\left[\hat{\rho}\right],
\end{equation}
is the reduced density matrix of the cluster.
In particular, the whole bath propagator is factorized into all possible cluster correlations as
\begin{equation}
{\mathcal{L}}=\prod_{\mathcal{C}\subseteq\{1,2,\cdots,N\}}\tilde{\mathcal{L}%
}_{\mathcal{C}}, \label{CCE_ENS_MULTI}%
\end{equation}
which assumes exactly the same form as Eq. (\ref{CCE_ENS_SINGLE}).

Now we discuss the convergence property of the ensemble CCE for
general bath operators. To focus on the effect of the interaction
strength, we consider the case that the initial bath density matrix
is factorizable as in Eq.~(\ref{factorized}). For a factorizable
ensemble, it can be readily proved that
$\ln\tilde{\mathcal{L}}_{\mathcal{C}}$ vanishes when the
interactions contained in operators $\hat{O}_{\mathcal{C}}^{(1)}$,
$\hat{O}_{\mathcal{C}}^{(2)}$, $\cdots$ cannot connect all the spins
in the group $\mathcal{C}$ into a linked cluster. Considering that
the operators $\hat{O}^{(j)}$ contains the $W$-spin interaction such
as
$\beta_{i_{1}i_{2}i_{3}}^{(j)}\hat{J}_{i_{1}}^{x}\hat{J}_{i_{2}}^{y}\hat{J}_{i_{3}}^{z}$
for $W=3$ (in the example of Sec. \ref{example}, $W=2$), in the
Taylor expansion of  $\ln\tilde{\mathcal{L}}_{\mathcal{C}}$ with
respect to the dimensionless coupling coefficients $\beta$'s, the
interaction coefficients $\{\beta_{i_{1}
,i_{2},\cdots,i_{W}}^{(j)}\}$ contained in each term must connect
all the spins in group $\mathcal{C}$ into a linked cluster. At least
$\left( M-1\right) /\left(  W-1\right) $ interaction coefficients
are needed to form a size-$M$ linked cluster. As a result,
\begin{equation}
\ln\tilde{\mathcal{L}}_{\mathcal{C}}=O\left(\beta^{\left( M-1\right)  /\left(  W-1\right)}\right),
\end{equation}
where $\beta$ is the typical magnitude of the coupling coefficients.
The number of size-$M$ clusters is $\sim Nq^{M-1}$.
So the total contribution to $\ln\mathcal{L}$ from all size-$M$ clusters is
\[
\sum_{\left\vert \mathcal{C}\right\vert =M}\ln\tilde{\mathcal{L}}%
_{\mathcal{C}}\sim N\left(  q^{W-1}\beta\right)  ^{(M-1)/(W-1)}.
\]
Therefore, the ensemble CCE converges for $q^{W-1}\beta<1$.
Interestingly, if the bath has certain initial correlations or entanglement
(such as a strongly correlated system at low temperature),
the convergence property would be determined by both the inverse temperature $\beta_T$
and the typical coupling constant $\beta$.

\subsection{Comparison with single-sample CCE}

The evaluation of the single-sample average
\begin{equation}
\mathcal{L}^{\mathcal{J}}=\left\langle \mathcal{J}\left\vert e^{i\hat{O}^{(1)}}e^{i\hat{O}^{(2)}}%
\cdots\right\vert \mathcal{J}\right\rangle \label{L_SINGLE_MULTI}%
\end{equation}
on a factorizable product state $\left\vert \mathcal{J}\right\rangle
=\otimes_{n}\left\vert j_{n}\right\rangle $ can be done as a special
case of the ensemble CCE by taking $\hat{\rho}=\left\vert
\mathcal{J}\right\rangle \left\langle \mathcal{J}\right\vert
=\otimes_{n}\hat{\rho}_{\{n\}}$ with $\hat{\rho
}_{\{n\}}\equiv\left\vert j_{n}\right\rangle \left\langle
j_{n}\right\vert$. This method is slightly different from the
previously developed single-sample CCE,\cite{Yang2008CCE} in
defining the cluster correlation for a cluster $\mathcal{C}$. Here
not only off-diagonal but also diagonal interaction terms involving
spins outside cluster $\mathcal{C}$ have been dropped. In contrast,
the single-sample CCE in Ref.~\onlinecite{Yang2008CCE} keeps all the
diagonal terms by replacing the spins outside cluster $\mathcal{C}$
with their mean-field values in the initial state $|{\mathcal
J}\rangle$. For example, the diagonal interaction $ \sum_{n\notin
{\mathcal C}}\gamma_{i,n}\hat{J}_i^z\hat{J}_n^z $ between a spin
$\hat{\mathbf{J}}_{i\in\mathcal{C}}$ inside cluster $\mathcal{C}$
and spins $\{\hat{\mathbf{J}}_{n\notin\mathcal{C}}\}$ outside
cluster $\mathcal{C}$ would be replaced with $ \hat{J}_i^z
\sum_{n\notin {\mathcal C}}\gamma_{i,n}\langle {\mathcal J}|
\hat{J}_n^z |{\mathcal J}\rangle $, which contributes a {\em static}
local mean-field for the $i$th spin. With such a static mean-field
treatment of the diagonal terms, the expansion is carried out with
respect to the most essential dynamics of the spin bath, namely, the
collective flip-flop of a cluster of spins that is responsible for
the {\em dynamical} local field fluctuation for the qubit. This
procedure, however, is not applicable to the ensemble CCE, for each
sample state $\left | \mathcal{J}\right \rangle$ from the ensemble
$\hat{\rho}$ will generate a different static mean field.

With the static local field fluctuation singled out, the cluster
correlation in the single-sample CCE accounts for the collective
dynamical local field fluctuation generated by off-diagonal
interactions. As a result, its magnitude in single-sample CCE is
determined by the magnitude of off-diagonal interactions, while in
ensemble CCE it is determined by the magnitude of all kinds of bath
interactions. As an example, consider pairwise interaction like
\[
\hat{O}\equiv\sum_{n}\alpha_{n}\hat{J}_{n}^{z}+\sum_{m<n}\left[
\beta^{\rm{nd}}
_{m,n}(\hat{J}_{m}^{+}\hat{J}_{n}^{-}+\hat{J}_{m}^{-}\hat{J}_{n}^{+}%
)+\beta^{\rm{d}}_{m,n}\hat{J}_{m}^{z}\hat{J}_{n}^{z}\right].
\]
Let $\beta_{\rm{nd}}$ ($\beta_{\rm{d}}$) be the typical magnitude of
the off-diagonal (diagonal) coupling coefficients
$\beta^{\rm{nd}}_{m,n}$ ($\beta^{\rm{d}}_{m,n}$) and $\beta$ be the
greater one of $\beta_{\rm{d}}$ and $\beta_{\rm{nd}}$. Then for a
size-$M$ cluster,
$\ln\mathcal{\tilde{L}}_{\mathcal{C}}=O(\beta^{M-1})$ in the
ensemble CCE, while $\ln\tilde{\mathcal L}^{\mathcal
J}_{\mathcal{C}}=O(\beta_{\rm{nd}}^{M-1})$ in the single-sample CCE.
The single-sample CCE would converge faster than the ensemble CCE.
Moreover, the number of clusters in ensemble CCE is greater than
that of the single-sample CCE. For example, if a spin
$\hat{\mathbf{J}}_{i}$ in a cluster interacts with others through
diagonal interaction only, then it generates no dynamical
fluctuations and hence the single-sample cluster correlation
vanishes, while the ensemble cluster correlation does not. For a
bath with a relatively large number of spins and a factorizable
initial state, the ensemble CCE result would be close to the
single-sample CCE using a random sampling of the initial bath state.
The single-sample CCE is recommended in such cases.
For a relatively small bath or a non-factorizable ensemble, the ensemble
CCE is desirable.

\section{Numerical check}

For a qubit-bath system described by a general pure dephasing Hamiltonian as
in Eq. (\ref{pure_dephasing_H}), the decoherence of the qubit under the pulse
control of the $n$th-order concatenated dynamical decoupling
\cite{Khodjasteh2005_PRL,Khodjasteh2007,Santos2006,Yao2007_RestoreCoherence,Witzel2007_PRBCDD}
is characterized by \cite{Liu2007_NJP}
\begin{equation}
\mathcal{L}_{n}\equiv\operatorname*{Tr}\left[\hat{\rho}\hat{U}_{n}^{(-)\dagger}%
U_{n}^{(+)}\right],\label{LN_ENS}%
\end{equation}
where $\hat{\rho}$ is the density matrix for the initial bath state and
$\hat{U}_{n}^{(\pm)}$ are recursively defined as
\[
\hat{U}_{j}^{(\pm)}\equiv\hat{U}_{j-1}^{(\mp)}\hat{U}_{j-1}^{(\pm)}%
\]
with $\hat{U}_{0}^{(\pm)}\equiv e^{-i\hat{H}^{(\pm)}T}$. For example,
free-induction decay, Hahn echo, and Carr-Purcell echo correspond to $n=0$, 1,
and $2$, respectively.

In this section, we consider an exactly solvable spin bath model (the
one-dimensional spin-1/2 XY model) and compare the qubit coherence
$\mathcal{L}_{n}$ from the ensemble CCE with the exact solutions by the
Jordan-Wigner transformation.\cite{Lieb1961,Huang2006} The $N$-spin bath
Hamiltonian conditioned on the qubit state $|\pm\rangle$ is
\begin{equation}
\hat{H}^{(\pm)}=\pm\sum_{n=1}^{N}\frac{z_{n}}{2}\hat{J}_{n}^{z}+\sum
_{n=1}^{N-1}\left(  B_{n}\pm\frac{b_{n}}{2}\right)  \left(  \hat{J}_{n+1}%
^{+}\hat{J}_{n}^{-}+\hat{J}_{n}^{+}\hat{J}_{n+1}^{-}\right)
,\label{HNPxymodel}%
\end{equation}
where $z_{n}$ denotes the qubit-bath spin interaction strength (simulating
the hyperfine interaction strength for electron-nuclear spin
systems), $B_{n}$ is the intrinsic bath interaction strength, and
$b_{n}$ is the interaction dependent on the qubit state. The bath is
assumed to be in a high-temperature thermal ensemble with
$\hat{\rho}\equiv\left(1/2\right)^{N}$. The qubit-bath interaction
coefficients $\{z_{n}\}$ are taken from a sinusoidal distribution
$z_{n}=z_{\mathrm{max}}\sin(n\pi/N)$ (referred to as
\textquotedblleft sinusoidal{\textquotedblright} chain) or randomly
selected from $[0,z_{\max}]$ (referred to as \textquotedblleft
random{\textquotedblright} chain). Hereafter $z_{\mathrm{max}}$ is
taken as the unit of energy. The spin-flip interaction
strengths $\{B_{n}\}$ and $\{b_{n}\}$ are randomly chosen from $[10^{-3}%
,2\times10^{-3}]$, corresponding to typical bath spin flip-flop time
$\tau_{\mathrm{sf}}\sim10^{3}$. The convergence of the ensemble CCE then
requires $B_{n}T,b_{n}T< 1$ or equivalently $T<\tau_{\mathrm{sf}}$.

\begin{figure}[ptb]
\includegraphics[width=\columnwidth]{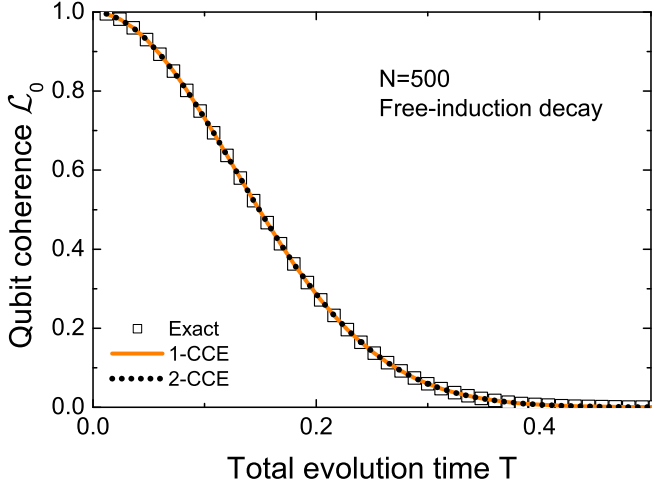}\caption{(Color online) Qubit
coherence in free-induction decay for a \textquotedblleft
sinusoidal{\textquotedblright} chain with $N=500$ spins: the exact
solution (empty squares) vs. the results from ensemble CCE truncated
to the first order (1-CCE, solid line) and the second order (2-CCE, dotted line).}%
\label{G_FID}%
\end{figure}

First we consider the simplest case, namely, the qubit coherence $\mathcal{L}%
_{0}=\operatorname*{Tr}\left[\hat{\rho}e^{i\hat{H}^{(-)}T}e^{-i\hat{H}^{(+)}T}\right]$ in
free-induction decay. The first-order truncation of the ensemble CCE gives
\begin{equation}
\mathcal{L}_{0}^{(1)}=\prod_{i}\cos\left(  \frac{z_{i}T}{2}\right),\label{L0_1CCE}%
\end{equation}
which is indeed the dephasing due to inhomogeneous broadening.
In the short-time limit ($z_{i}T\ll 1$), Eq.~(\ref{L0_1CCE}) becomes
$\mathcal{L}_{0}^{(1)}\approx e^{-\Gamma^{2}T^{2}/2}$, where
\[
\Gamma\equiv\sqrt{\frac{1}{4}\sum_{i}z_{i}^{2}}=\sqrt{\left\langle (\hat
{h}^{z})^{2}\right\rangle -\left\langle \hat{h}^{z}\right\rangle ^{2}}
\sim \sqrt{N}\left|z_i\right|,
\]
is the variance of the ``Overhauser'' field $\hat{h}^{z}\equiv\sum_{n}z_{n}\hat
{J}_{n}^{z}$ of the spin bath. As evidenced by
the good agreement between the 1-CCE and the exact result in Fig.~\ref{G_FID},
the ensemble free-induction decay is dominated by the inhomogeneous
broadening, which leads to rapid decoherence within a time scale much shorter
than the inverse qubit-bath interaction strength $z_{n}^{-1}\sim 1$. The corrections due to
spin-spin interactions, which could show up on a time scale comparable with
the inverse interaction strength ($1/B_{n},1/b_{n}\sim10^{3}$) and dominates
the single-sample decoherence, is negligible during the ensemble
free-induction decay process.

\begin{figure}[ptb]
\includegraphics[width=\columnwidth]{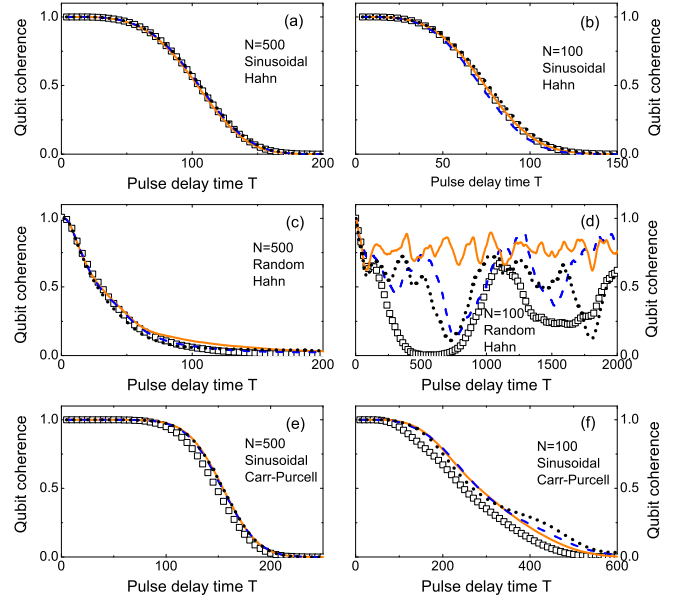}\caption{(Color
online) Qubit coherence in Hahn echo ($n$=1) and Carr-Purcell echo ($n=2$) for
different baths. The exact ensemble coherence $\mathcal{L}_{n}$ (empty
squares) vs. the magnitude $|{\mathcal L}_{n}^{(\mathcal{J})}|$ of the exact
single-sample coherence for three randomly chosen bath states $|\mathcal{J}%
\rangle$, denoted by the solid, dashed, and dotted lines, respectively.}%
\label{G_ENS_SAMPLE}%
\end{figure}

To highlight the role of spin-spin interaction, we consider the
qubit coherence $\mathcal{L}_{n}$ $(n=1,2,\cdots)$ in spin echo or
higher order concatenated control where  the inhomogeneous
broadening is eliminated and, consequently, the first-order
truncation of the ensemble CCE gives no decay:
$\mathcal{L}_{1}^{(1)}=1$. When the random phase factor leading to
the inhomogeneous broadening is eliminated (as in the echo signals),
the ensemble-averaged qubit coherence would be close to that
averaged on a randomly sampled bath state, if the bath is relatively
large. This is shown in Figs.~\ref{G_ENS_SAMPLE}~(a) and
\ref{G_ENS_SAMPLE}~(b) for a {\em sinusoidal} chain: The
single-sample coherences for three randomly chosen bath states
$\left\vert \mathcal{J}\right\rangle $ agree very well with the
ensemble coherence.

For a small spin bath, or for a {\em random} chain where the qubit
decoherence is caused by the dynamics of a few small clusters, the
qubit decoherence would depend sensitively on the choice of the
initial bath state and hence the ensemble average would deviate
significantly from the qubit coherence averaged on any specific
sample of the initial state. This is clearly seen in
Figs.~\ref{G_ENS_SAMPLE}~(c) and \ref{G_ENS_SAMPLE}~(d) for a {\em
random} chain. Even for a relatively large {\em sinusoidal} chain,
the difference between single-sample decoherence and the ensemble
average is noticeable when higher-order dynamical decoupling (e.g.,
Carr-Purcell echo) is applied [see Figs. \ref{G_ENS_SAMPLE}(e) and
\ref{G_ENS_SAMPLE}(f)]. This is because that under the higher-order
control, the clusters responsible for the qubit decoherence grow
larger and larger as the effects of smaller clusters are suppressed,
and the specificity of the initial state of larger clusters is more
important than that of smaller ones.

\begin{figure}[ptb]
\includegraphics[width=\columnwidth]{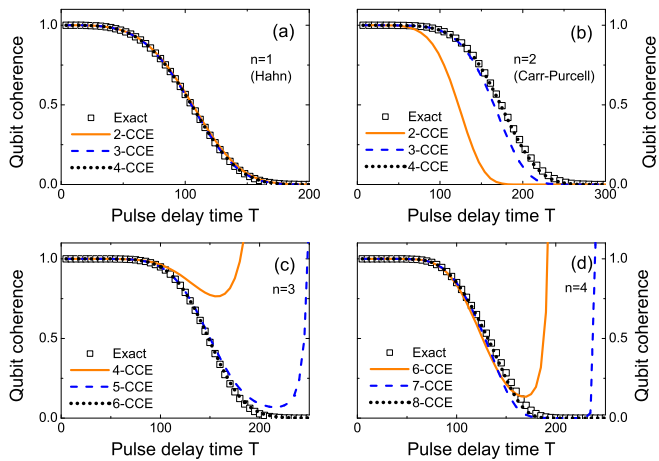}\caption{(Color
online) Ensemble qubit coherence in $n$th-order concatenated
dynamical decoupling for a {\em sinusoidal} chain with $N=500$ spins: the exact
ensemble solutions $($empty squares$)$ vs. the results from the
ensemble CCE truncated
to different orders.}%
\label{G_Convergence_CDD}%
\end{figure}

Figure \ref{G_Convergence_CDD} compares the results from the
ensemble CCE with the exact solutions for ensemble coherence under
the control of concatenated dynamical decoupling of different
orders. For the Hahn echo shown in Fig. \ref{G_Convergence_CDD}(a),
the second order truncation of the ensemble CCE already agrees with
the exact solution very well, indicating that the decoherence is
dominated by spin-pair dynamics. For Carr-Purcell echo in Fig.
\ref{G_Convergence_CDD}(b), however, the second-order truncation
becomes insufficient, as the leading order contributing from
spin-pair dynamics has been eliminated and the correlated dynamics
of larger clusters becomes important \cite{Witzel2007_PRBCDD}. Hence
a fourth-order truncation of the ensemble CCE is required to
reproduce the exact solution. For concatenated dynamical decoupling
of successively higher orders $n$, the leading contributions from
successively larger clusters have been eliminated. As a result,
successively higher-order truncations of the ensemble CCE are
required to reproduce the exact results: 6-CCE for $n=3$ in Fig.
\ref{G_Convergence_CDD}(c) and 8-CCE for $n=4$ in Fig.
\ref{G_Convergence_CDD}(d).

\section{Conclusion}

The decoherence of a qubit in a spin bath is essentially determined by the
many-body bath evolution, starting either from a pure or an ensemble state.
As an extension of the previously developed single-sample
cluster-correlation expansion\cite{Yang2008CCE} that addresses the spin
bath dynamics starting from a noninteracting pure state, we have developed an
ensemble CCE theory to solve the spin bath dynamics starting from an
arbitrary ensemble state. In this approach, the ensemble propagator is
factorized into the product of all possible cluster correlations, each term
accounting for the authentic (non-factorizable) collective excitation of a
group of bath spins. For a finite-time evolution as in the qubit decoherence
problem, convergent results can be obtained by truncating the ensemble CCE by
keeping cluster correlations up to a certain size, as has been checked using
an exactly solvable spin chain model (the one-dimensional spin-1/2 XY model).
For the convergence property (for a factorizable initial state),
the ensemble CCE is determined by the typical strength
of the bath spin interaction, being either diagonal or off-diagonal,
while the single-sample CCE is determined only by the off-diagonal interaction
coefficients. The ensemble CCE can be applied to baths with
non-factorizable initial state as well.

\acknowledgments This work was supported by Hong Kong GRF Project 401906.


\begin{thebibliography}{32}
\expandafter\ifx\csname
natexlab\endcsname\relax\def\natexlab#1{#1}\fi
\expandafter\ifx\csname bibnamefont\endcsname\relax
  \def\bibnamefont#1{#1}\fi
\expandafter\ifx\csname bibfnamefont\endcsname\relax
  \def\bibfnamefont#1{#1}\fi
\expandafter\ifx\csname citenamefont\endcsname\relax
  \def\citenamefont#1{#1}\fi
\expandafter\ifx\csname url\endcsname\relax
  \def\url#1{\texttt{#1}}\fi
\expandafter\ifx\csname urlprefix\endcsname\relax\def\urlprefix{URL
}\fi \providecommand{\bibinfo}[2]{#2}
\providecommand{\eprint}[2][]{\url{#2}}

\bibitem[{\citenamefont{Prokof'ev and Stamp}(2000)}]{Prokofev2000_RPP}
\bibinfo{author}{\bibfnamefont{N.~V.} \bibnamefont{Prokof'ev}}
  \bibnamefont{and} \bibinfo{author}{\bibfnamefont{P.~C.~E.}
  \bibnamefont{Stamp}}, \bibinfo{journal}{Rep. Prog. Phys.}
  \textbf{\bibinfo{volume}{63}}, \bibinfo{pages}{669} (\bibinfo{year}{2000}).

\bibitem[{\citenamefont{Kubo}(1954)}]{Kubo1954_JPSJ}
\bibinfo{author}{\bibfnamefont{R.}~\bibnamefont{Kubo}}, \bibinfo{journal}{J.
  Phys. Soc. Jpn.} \textbf{\bibinfo{volume}{9}}, \bibinfo{pages}{935}
  (\bibinfo{year}{1954}).

\bibitem[{\citenamefont{Pines and Slichter}(1955)}]{Pines1955}
\bibinfo{author}{\bibfnamefont{D.}~\bibnamefont{Pines}} \bibnamefont{and}
  \bibinfo{author}{\bibfnamefont{C.~P.} \bibnamefont{Slichter}},
  \bibinfo{journal}{Phys. Rev.} \textbf{\bibinfo{volume}{100}},
  \bibinfo{pages}{1014} (\bibinfo{year}{1955}).

\bibitem[{\citenamefont{Feher and Gere}(1959)}]{Feher1959}
\bibinfo{author}{\bibfnamefont{G.}~\bibnamefont{Feher}} \bibnamefont{and}
  \bibinfo{author}{\bibfnamefont{E.~A.} \bibnamefont{Gere}},
  \bibinfo{journal}{Phys. Rev.} \textbf{\bibinfo{volume}{114}},
  \bibinfo{pages}{1245} (\bibinfo{year}{1959}).

\bibitem[{\citenamefont{Klauder and Anderson}(1962)}]{Klauder1962}
\bibinfo{author}{\bibfnamefont{J.~R.} \bibnamefont{Klauder}} \bibnamefont{and}
  \bibinfo{author}{\bibfnamefont{P.~W.} \bibnamefont{Anderson}},
  \bibinfo{journal}{Phys. Rev.} \textbf{\bibinfo{volume}{125}},
  \bibinfo{pages}{912 } (\bibinfo{year}{1962}).

\bibitem[{\citenamefont{Fujisawa et~al.}(2002)\citenamefont{Fujisawa, Austing,
  Tokura, Hirayama, and Tarucha}}]{Fujisawa2002}
\bibinfo{author}{\bibfnamefont{T.}~\bibnamefont{Fujisawa}},
  \bibinfo{author}{\bibfnamefont{D.~G.} \bibnamefont{Austing}},
  \bibinfo{author}{\bibfnamefont{Y.}~\bibnamefont{Tokura}},
  \bibinfo{author}{\bibfnamefont{Y.}~\bibnamefont{Hirayama}}, \bibnamefont{and}
  \bibinfo{author}{\bibfnamefont{S.}~\bibnamefont{Tarucha}},
  \bibinfo{journal}{Nature} \textbf{\bibinfo{volume}{419}}, \bibinfo{pages}{278
  } (\bibinfo{year}{2002}).

\bibitem[{\citenamefont{Elzerman et~al.}(2004)\citenamefont{Elzerman, Hanson,
  Willems~van Beveren, Witkamp, Vandersypen, and
  Kouwenhoven}}]{Elzerman2004_Nature}
\bibinfo{author}{\bibfnamefont{J.~M.} \bibnamefont{Elzerman}},
  \bibinfo{author}{\bibfnamefont{R.}~\bibnamefont{Hanson}},
  \bibinfo{author}{\bibfnamefont{L.~H.} \bibnamefont{Willems~van Beveren}},
  \bibinfo{author}{\bibfnamefont{B.}~\bibnamefont{Witkamp}},
  \bibinfo{author}{\bibfnamefont{L.~M.~K.} \bibnamefont{Vandersypen}},
  \bibnamefont{and} \bibinfo{author}{\bibfnamefont{L.~P.}
  \bibnamefont{Kouwenhoven}}, \bibinfo{journal}{Nature}
  \textbf{\bibinfo{volume}{430}}, \bibinfo{pages}{431} (\bibinfo{year}{2004}).

\bibitem[{\citenamefont{Kroutvar et~al.}(2004)\citenamefont{Kroutvar, Ducommun,
  Heiss, Bichler, Schuh, Abstreiter, and Finley}}]{Kroutvar2004_Nature}
\bibinfo{author}{\bibfnamefont{M.}~\bibnamefont{Kroutvar}},
  \bibinfo{author}{\bibfnamefont{Y.}~\bibnamefont{Ducommun}},
  \bibinfo{author}{\bibfnamefont{D.}~\bibnamefont{Heiss}},
  \bibinfo{author}{\bibfnamefont{M.}~\bibnamefont{Bichler}},
  \bibinfo{author}{\bibfnamefont{D.}~\bibnamefont{Schuh}},
  \bibinfo{author}{\bibfnamefont{G.}~\bibnamefont{Abstreiter}},
  \bibnamefont{and} \bibinfo{author}{\bibfnamefont{J.~J.}
  \bibnamefont{Finley}}, \bibinfo{journal}{Nature}
  \textbf{\bibinfo{volume}{432}}, \bibinfo{pages}{81} (\bibinfo{year}{2004}).

\bibitem[{\citenamefont{Khaetskii and Nazarov}(2000)}]{Khaetskii2000}
\bibinfo{author}{\bibfnamefont{A.~V.} \bibnamefont{Khaetskii}}
  \bibnamefont{and} \bibinfo{author}{\bibfnamefont{Y.}~\bibnamefont{Nazarov}},
  \bibinfo{journal}{Phys. Rev. B} \textbf{\bibinfo{volume}{61}},
  \bibinfo{pages}{12639} (\bibinfo{year}{2000}).

\bibitem[{\citenamefont{Woods et~al.}(2002)\citenamefont{Woods, Reinecke, and
  Lyanda-Geller}}]{Woods2002}
\bibinfo{author}{\bibfnamefont{L.~M.} \bibnamefont{Woods}},
  \bibinfo{author}{\bibfnamefont{T.~L.} \bibnamefont{Reinecke}},
  \bibnamefont{and}
  \bibinfo{author}{\bibfnamefont{Y.}~\bibnamefont{Lyanda-Geller}},
  \bibinfo{journal}{Phys. Rev. B} \textbf{\bibinfo{volume}{66}},
  \bibinfo{pages}{161318(R)} (\bibinfo{year}{2002}).

\bibitem[{\citenamefont{Golovach et~al.}(2004)\citenamefont{Golovach,
  Khaetskii, and Loss}}]{Golovach2004}
\bibinfo{author}{\bibfnamefont{V.~N.} \bibnamefont{Golovach}},
  \bibinfo{author}{\bibfnamefont{A.}~\bibnamefont{Khaetskii}},
  \bibnamefont{and} \bibinfo{author}{\bibfnamefont{D.}~\bibnamefont{Loss}},
  \bibinfo{journal}{Phys. Rev. Lett.} \textbf{\bibinfo{volume}{93}},
  \bibinfo{pages}{016601} (\bibinfo{year}{2004}).

\bibitem[{\citenamefont{Semenov and Kim}(2004)}]{Semenov2004}
\bibinfo{author}{\bibfnamefont{Y.~G.} \bibnamefont{Semenov}} \bibnamefont{and}
  \bibinfo{author}{\bibfnamefont{K.~W.} \bibnamefont{Kim}},
  \bibinfo{journal}{Phys. Rev. Lett.} \textbf{\bibinfo{volume}{92}},
  \bibinfo{pages}{026601} (\bibinfo{year}{2004}).

\bibitem[{\citenamefont{Witzel et~al.}(2005)\citenamefont{Witzel, de~Sousa, and
  Das~Sarma}}]{Witzel2005}
\bibinfo{author}{\bibfnamefont{W.~M.} \bibnamefont{Witzel}},
  \bibinfo{author}{\bibfnamefont{R.}~\bibnamefont{de~Sousa}}, \bibnamefont{and}
  \bibinfo{author}{\bibfnamefont{S.}~\bibnamefont{Das~Sarma}},
  \bibinfo{journal}{Phys. Rev. B} \textbf{\bibinfo{volume}{72}},
  \bibinfo{pages}{161306(R)} (\bibinfo{year}{2005}).

\bibitem[{\citenamefont{Witzel and Das~Sarma}(2006)}]{Witzel2006}
\bibinfo{author}{\bibfnamefont{W.~M.} \bibnamefont{Witzel}} \bibnamefont{and}
  \bibinfo{author}{\bibfnamefont{S.}~\bibnamefont{Das~Sarma}},
  \bibinfo{journal}{Phys. Rev. B} \textbf{\bibinfo{volume}{74}},
  \bibinfo{pages}{035322} (\bibinfo{year}{2006}).

\bibitem[{\citenamefont{Witzel and Das~Sarma}(2007{\natexlab{a}})}]{Witzel2007}
\bibinfo{author}{\bibfnamefont{W.~M.} \bibnamefont{Witzel}} \bibnamefont{and}
  \bibinfo{author}{\bibfnamefont{S.}~\bibnamefont{Das~Sarma}},
  \bibinfo{journal}{Phys. Rev. Lett.} \textbf{\bibinfo{volume}{98}},
  \bibinfo{pages}{077601} (\bibinfo{year}{2007}{\natexlab{a}}).

\bibitem[{\citenamefont{Witzel and
  Das~Sarma}(2007{\natexlab{b}})}]{Witzel2007_PRB}
\bibinfo{author}{\bibfnamefont{W.~M.} \bibnamefont{Witzel}} \bibnamefont{and}
  \bibinfo{author}{\bibfnamefont{S.}~\bibnamefont{Das~Sarma}},
  \bibinfo{journal}{Phys. Rev. B} \textbf{\bibinfo{volume}{76}},
  \bibinfo{pages}{045218} (\bibinfo{year}{2007}{\natexlab{b}}).

\bibitem[{\citenamefont{Witzel and
  Das~Sarma}(2007{\natexlab{c}})}]{Witzel2007_PRBCDD}
\bibinfo{author}{\bibfnamefont{W.~M.} \bibnamefont{Witzel}} \bibnamefont{and}
  \bibinfo{author}{\bibfnamefont{S.}~\bibnamefont{Das~Sarma}},
  \bibinfo{journal}{Phys. Rev. B} \textbf{\bibinfo{volume}{76}},
  \bibinfo{pages}{241303(R)} (\bibinfo{year}{2007}{\natexlab{c}}).

\bibitem[{\citenamefont{Yao et~al.}(2006)\citenamefont{Yao, Liu, and
  Sham}}]{Yao2006_PRB}
\bibinfo{author}{\bibfnamefont{W.}~\bibnamefont{Yao}},
  \bibinfo{author}{\bibfnamefont{R.~B.} \bibnamefont{Liu}}, \bibnamefont{and}
  \bibinfo{author}{\bibfnamefont{L.~J.} \bibnamefont{Sham}},
  \bibinfo{journal}{Phys. Rev. B} \textbf{\bibinfo{volume}{74}},
  \bibinfo{pages}{195301} (\bibinfo{year}{2006}).

\bibitem[{\citenamefont{Yao et~al.}(2007)\citenamefont{Yao, Liu, and
  Sham}}]{Yao2007_RestoreCoherence}
\bibinfo{author}{\bibfnamefont{W.}~\bibnamefont{Yao}},
  \bibinfo{author}{\bibfnamefont{R.~B.} \bibnamefont{Liu}}, \bibnamefont{and}
  \bibinfo{author}{\bibfnamefont{L.~J.} \bibnamefont{Sham}},
  \bibinfo{journal}{Phys. Rev. Lett.} \textbf{\bibinfo{volume}{98}},
  \bibinfo{pages}{077602} (\bibinfo{year}{2007}).

\bibitem[{\citenamefont{Liu et~al.}(2007)\citenamefont{Liu, Yao, and
  Sham}}]{Liu2007_NJP}
\bibinfo{author}{\bibfnamefont{R.~B.} \bibnamefont{Liu}},
  \bibinfo{author}{\bibfnamefont{W.}~\bibnamefont{Yao}}, \bibnamefont{and}
  \bibinfo{author}{\bibfnamefont{L.~J.} \bibnamefont{Sham}},
  \bibinfo{journal}{New J. Phys.} \textbf{\bibinfo{volume}{9}},
  \bibinfo{pages}{226} (\bibinfo{year}{2007}).

\bibitem[{\citenamefont{Deng and Hu}(2006)}]{Deng2006}
\bibinfo{author}{\bibfnamefont{C.}~\bibnamefont{Deng}} \bibnamefont{and}
  \bibinfo{author}{\bibfnamefont{X.}~\bibnamefont{Hu}}, \bibinfo{journal}{Phys.
  Rev. B} \textbf{\bibinfo{volume}{73}}, \bibinfo{pages}{241303(R)}
  (\bibinfo{year}{2006}).

\bibitem[{\citenamefont{Saikin et~al.}(2007)\citenamefont{Saikin, Yao, and
  Sham}}]{Saikin2007}
\bibinfo{author}{\bibfnamefont{S.~K.} \bibnamefont{Saikin}},
  \bibinfo{author}{\bibfnamefont{W.}~\bibnamefont{Yao}}, \bibnamefont{and}
  \bibinfo{author}{\bibfnamefont{L.~J.} \bibnamefont{Sham}},
  \bibinfo{journal}{Phys. Rev. B} \textbf{\bibinfo{volume}{75}},
  \bibinfo{pages}{125314} (\bibinfo{year}{2007}).

\bibitem[{\citenamefont{Yang and Liu}(2008{\natexlab{a}})}]{YangDQD2008}
\bibinfo{author}{\bibfnamefont{W.}~\bibnamefont{Yang}} \bibnamefont{and}
  \bibinfo{author}{\bibfnamefont{R.~B.} \bibnamefont{Liu}},
  \bibinfo{journal}{Phys. Rev. B} \textbf{\bibinfo{volume}{77}},
  \bibinfo{pages}{085302} (\bibinfo{year}{2008}{\natexlab{a}}).

\bibitem[{\citenamefont{Yang and Liu}(2008{\natexlab{b}})}]{Yang2008CCE}
\bibinfo{author}{\bibfnamefont{W.}~\bibnamefont{Yang}} \bibnamefont{and}
  \bibinfo{author}{\bibfnamefont{R.~B.} \bibnamefont{Liu}},
  \bibinfo{journal}{Phys. Rev. B} \textbf{\bibinfo{volume}{78}},
  \bibinfo{pages}{085315} (\bibinfo{year}{2008}{\natexlab{b}}).

\bibitem[{\citenamefont{Loss and DiVincenzo}(1998)}]{Loss1998}
\bibinfo{author}{\bibfnamefont{D.}~\bibnamefont{Loss}} \bibnamefont{and}
  \bibinfo{author}{\bibfnamefont{D.~P.} \bibnamefont{DiVincenzo}},
  \bibinfo{journal}{Phys. Rev. A} \textbf{\bibinfo{volume}{57}},
  \bibinfo{pages}{120 } (\bibinfo{year}{1998}).

\bibitem[{\citenamefont{Imamo\={g}lu et~al.}(1999)\citenamefont{Imamo\={g}lu,
  Awschalom, Burkard, DiVincenzo, Loss, Sherwin, and Small}}]{Imamoglu1999}
\bibinfo{author}{\bibfnamefont{A.}~\bibnamefont{Imamo\={g}lu}},
  \bibinfo{author}{\bibfnamefont{D.~D.} \bibnamefont{Awschalom}},
  \bibinfo{author}{\bibfnamefont{G.}~\bibnamefont{Burkard}},
  \bibinfo{author}{\bibfnamefont{D.~P.} \bibnamefont{DiVincenzo}},
  \bibinfo{author}{\bibfnamefont{D.}~\bibnamefont{Loss}},
  \bibinfo{author}{\bibfnamefont{M.}~\bibnamefont{Sherwin}}, \bibnamefont{and}
  \bibinfo{author}{\bibfnamefont{A.}~\bibnamefont{Small}},
  \bibinfo{journal}{Phys. Rev. Lett.} \textbf{\bibinfo{volume}{83}},
  \bibinfo{pages}{4204 } (\bibinfo{year}{1999}).

\bibitem[{\citenamefont{Awschalom et~al.}(2002)\citenamefont{Awschalom, Loss,
  and Samarth}}]{Awschalom2002_book}
\bibinfo{editor}{\bibfnamefont{D.~D.} \bibnamefont{Awschalom}},
  \bibinfo{editor}{\bibfnamefont{D.}~\bibnamefont{Loss}}, \bibnamefont{and}
  \bibinfo{editor}{\bibfnamefont{N.}~\bibnamefont{Samarth}}, eds.,
  \emph{\bibinfo{title}{Semiconductor Spintronics and Quantum Computation}}
  (\bibinfo{publisher}{Springer, New York}, \bibinfo{year}{2002}).

\bibitem[{\citenamefont{Khodjasteh and Lidar}(2005)}]{Khodjasteh2005_PRL}
\bibinfo{author}{\bibfnamefont{K.}~\bibnamefont{Khodjasteh}} \bibnamefont{and}
  \bibinfo{author}{\bibfnamefont{D.~A.} \bibnamefont{Lidar}},
  \bibinfo{journal}{Phys. Rev. Lett.} \textbf{\bibinfo{volume}{95}},
  \bibinfo{pages}{180501} (\bibinfo{year}{2005}).

\bibitem[{\citenamefont{Khodjasteh and Lidar}(2007)}]{Khodjasteh2007}
\bibinfo{author}{\bibfnamefont{K.}~\bibnamefont{Khodjasteh}} \bibnamefont{and}
  \bibinfo{author}{\bibfnamefont{D.~A.} \bibnamefont{Lidar}},
  \bibinfo{journal}{Phys. Rev. A} \textbf{\bibinfo{volume}{75}},
  \bibinfo{pages}{062310} (\bibinfo{year}{2007}).

\bibitem[{\citenamefont{Santos and Viola}(2006)}]{Santos2006}
\bibinfo{author}{\bibfnamefont{L.~F.} \bibnamefont{Santos}} \bibnamefont{and}
  \bibinfo{author}{\bibfnamefont{L.}~\bibnamefont{Viola}},
  \bibinfo{journal}{Phys. Rev. Lett.} \textbf{\bibinfo{volume}{97}},
  \bibinfo{pages}{150501} (\bibinfo{year}{2006}).

\bibitem[{\citenamefont{Lieb et~al.}(1961)\citenamefont{Lieb, Schultz, and
  Mattis}}]{Lieb1961}
\bibinfo{author}{\bibfnamefont{E.}~\bibnamefont{Lieb}},
  \bibinfo{author}{\bibfnamefont{T.}~\bibnamefont{Schultz}}, \bibnamefont{and}
  \bibinfo{author}{\bibfnamefont{D.}~\bibnamefont{Mattis}},
  \bibinfo{journal}{Ann. Phys.} \textbf{\bibinfo{volume}{16}},
  \bibinfo{pages}{407} (\bibinfo{year}{1961}).

\bibitem[{\citenamefont{Huang et~al.}(2006)\citenamefont{Huang, Sadiek, and
  Kais}}]{Huang2006}
\bibinfo{author}{\bibfnamefont{Z.}~\bibnamefont{Huang}},
  \bibinfo{author}{\bibfnamefont{G.}~\bibnamefont{Sadiek}}, \bibnamefont{and}
  \bibinfo{author}{\bibfnamefont{S.}~\bibnamefont{Kais}}, \bibinfo{journal}{J.
  Chem. Phys.} \textbf{\bibinfo{volume}{124}}, \bibinfo{pages}{144513}
  (\bibinfo{year}{2006}).

\end{thebibliography}

\end{document}